# Macroscopic and Direct Light Propulsion of Bulk Graphene Material

Tengfei Zhang[1,†], Huicong Chang[1,†], Yingpeng Wu[1,†], Peishuang Xiao[1], Ningbo Yi[1], Yanhong Lu[1], Yanfeng Ma[1], Yi Huang[1], Kai Zhao[1], Xiao-Qing Yan[2], Zhi-Bo Liu[2], Jian-Guo Tian[2], Yongsheng Chen[1,*]

[1]Key Laboratory of Functional Polymer Materials and Center for Nanoscale Science and Technology, Collaborative Innovation Center of Chemical Science and Engineering, Institute of Polymer Chemistry, College of Chemistry, Nankai University, Tianjin 300071, China.
[2]Key Laboratory of Weak Light Nonlinear Photonics, Ministry of Education, Teda Applied Physics School and School of Physics, Nankai University, Tianjin 300457, China.

[†]These authors contributed equally to this work.

*Correspondence to: yschen99@nankai.edu.cn

Using beams of light, scientists have been able to trap[1], move[2], levitate[3] and even pull[4] small objects at the microscopic scale, such as atoms and molecules, living cells and viruses, micro/nanoscopic particles, and also nano/micron-sized graphene sheets[5-7] on a small spatial scale, typically hundreds of microns[8]. Some efforts for enlarged optical manipulation distance by harnessing strong thermal forces[9] and robust manipulation of airborne micro-objects photophoretically with bottle beam[10] are also reported. Motion and rotation of millimeter-sized graphite disk by photoirradiation were realized when the graphite was magnetically levitated[11]. If aforementioned optical operations could be achieved for large objects at macroscopic spatial scale, significant applications such as the long-sought direct optical manipulation of macroscale objects including even the proposed solar sail and space transportation through laser or beam-powered propulsion could be realized. To acquire the required energy and momentum for propulsion, there have been two main proposed mechanisms, which are using laser to superheat propellant (or air) that then provides propulsion like conventional rockets[4,12,13] or obtaining propulsion directly from light pressure (radiation pressure) acting on a light sail structure (such as IKAROS spacecraft)[14,15].

It has been an important challenge to realize the intrinsic properties of single layer graphene in a bulk material since the re-stacking of graphene sheets diminishes most of those properties including that in electronic, photonic and even mechanical aspects. In this work, we show that if graphene sheets are assembled in the proper format in the bulk state, the resulted bulk material can not only retain the intrinsic properties of individual graphene sheets, but also allow their manifestation on a macroscopic scale. Here, we present the directly light-induced macroscopic propulsion and rotation of bulk graphene sponge material which was centimeter sized and milligram weight. The mechanism behind this novel phenomenon is believed to be an efficient light-induced ejected electron emission process, following an Auger-like path due to both the unique band structure of graphene and its macroscopic morphology of this unique material. The force generated from such a process/mechanism is much larger than the force generated directly from the conventional light pressure, which is much smaller than the force required to propel the samples. A series of control experiments were further carried out, which also excludes the laser beam ablation mechanism. The efficient light absorption of graphene[16,17] and easily achievable reverse saturation state[18,19], combined with the unique and limited hot electron relaxing mechanisms and channels[17,20], all due to the unique band structure of graphene, collectively make this bulk graphene material capable of efficiently emitting energetic electrons while it is under light illumination so that the net momentum generated by the ejected electrons can propel the bulk graphene sponge according to Newton's laws of motion.

The graphene sponge was synthesized using a modified method reported earlier[21] followed by a high temperature annealing in an inert environment, and the detailed procedure is described in the Supplementary Information (SI), with additional optical images of different sizes of the bulk material (Supplementary Fig. 1). The material exhibits a conductivity of ~0.5 S m$^{-1}$ with a density of ~1 mg mL$^{-1}$.

When cutting the graphene sponge by laser in air, we accidently observed the laser-induced actuation by naked eyes, which contrasts sharply with the earlier reported microscopic levitation or movement of micro objects due to light pressure[3,8]. To avoid the likely intervention of air, further systematic studies were carried out in vacuum environment (from $6.8 \times 10^{-4}$ to $5 \times 10^{-6}$ Torr) entirely to rule out (minimize) the possibility of heated air disturbance and to avoid the local combustion of graphene sponge due to the presence of oxygen. The setup for the experiments was shown in Fig. 1 for both the direct light-induced propulsion (Fig. 1a) and rotation (Fig. 1b) of our bulk graphene samples. To avoid the impact from friction, electrostatic attraction and also collision between the sample and the tube in light-induced propulsion, light-induced rotation for a quantitative relationship and mechanism investigation was carried out with the apparatus shown in Fig. 1b, where the graphene sponge was cut into a cuboid and a glass capillary (or metal wire) acting as an axis to penetrate through the center of the sample.

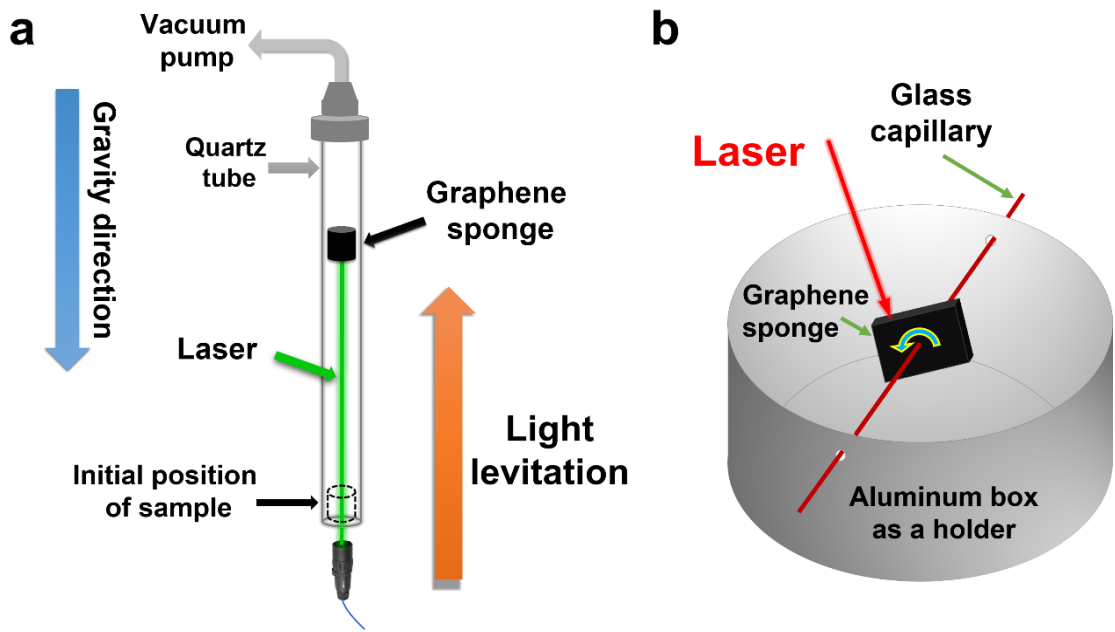

**Figure 1 | Measurement apparatuses and schematics of light-induced propulsion and rotation of graphene sponge. a**, Schematic of the graphene sponge being propelled vertical upwardly with underneath laser illumination. **b**, Apparatus of light-induced rotation of graphene sponge under laser illumination.

**Light-induced horizontal and vertical propulsion**

Firstly, as shown in Supplementary Video 1, macroscopic graphene objects (with centimeter scale size) could be pushed away immediately when the laser beam was applied, and lasers with different wavelengths (450, 532 and 650 nm) gave the same phenomenon. More surprisingly, when the graphene objects were put at the bottom of a vertical vacuum tube, direct and instant optical vertical upwardly propulsion to sub-meter height (due to the limit of the vacuum facilities) was observed (Figs. 2a and 2b, Supplementary Video 2) when the laser beam was shined underneath the sample. In Fig. 2a, when lasers with the same power density but different wavelengths were used, at the same moment after the same sample was illuminated, higher propulsion height was observed when lasers with shorter wavelength were used (Supplementary Video 2). As demonstrated in Fig. 2b (Supplementary Fig. 2), the propulsion height increased with the increasing laser power density if the laser wavelength was fixed no matter what laser wavelength was used. Similar dependence was also observed when the sample was placed in a horizontal vacuum tube (not shown). With the results above and the well-known band structure of graphene which in principle allows the absorption of all wavelengths of light[16,17], a simulated sunlight generated by a Xenon lamp was used as the light source for same test. Strikingly, similar direct sunlight-induced horizontal and vertical propulsion was achieved (Supplementary Video 3). Furthermore, by varying the distance between light source and sample to obtain light of different intensity, the vertical propulsion height varied accordingly (Supplementary Fig. 3). Most strikingly, by using the natural sunlight on a sunny day with a Fresnel lens for focusing, a similar optical response was observed (Supplementary Video 4).

**Rotation with different light wavelengths and intensities**

As mentioned above, the propulsion heights and speeds change with the light intensity and wavelength for a given sample. But to avoid other factors such as friction, electrostatic attraction and also collision between the sample and the tube in light-induced propulsion, a home-made device shown in Fig. 1b was used to obtain such a quantitative relationship for the laser-induced rotation (Supplementary Video 5) with different light wavelengths and power densities. The results for rotation speed versus laser power density/wavelength are summarized in Fig. 2c. Since the rotational kinetic energy $E$ is proportional to the square of rotation speed $r$ and sample mass $m$ ($E \propto mr^2$), the square of rotation speed was then plotted with laser wavelength and laser power density (detailed discussion in SI). Indeed, as shown in Fig. 2d, we found that for the same sample, the square of rotation speed (rotational kinetic energy) increased linearly with increasing laser power density at constant wavelength for all the laser beam tests (450, 532, 650 nm) and this dependence holds independently of the size and mass of the samples (Fig. 2e, Supplementary Fig. 4). Similarly, at a given laser power density, lasers with shorter wavelength (higher frequency and photon energy) gave a larger value of the square of the rotation speed (Fig. 2d), following the similar linear relationship which was demonstrated more clearly in Fig. 2f.

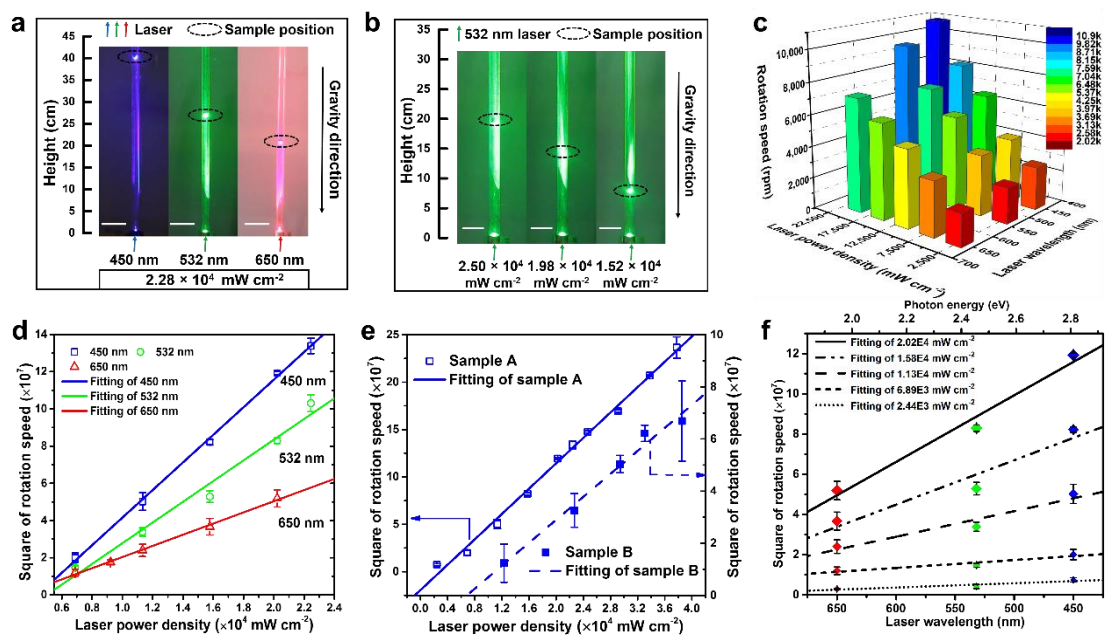

**Figure 2 | Relationships between laser-induced propulsion/rotation of graphene sponge and laser wavelength/power density. a**, Different vertical propulsion height of the same sample over the same time (2 s) and the same power density but different wavelengths (scale bars, 5 cm). **b**, Different vertical propulsion height of the same sample over the same time (1 s) and the same wavelength but different power densities

(scale bars, 5 cm). **c**, The 3D histogram showed the rotation speed of graphene sponge sample had a distinct positive correlation with power density and frequency of the laser used. **d**, The square of rotation speed increased linearly with laser power density; lasers with different wavelengths gave similar results. **e**, The linear relationship of laser power density and square of rotation speed for different samples (laser wavelength, 450 nm). **f**, For a certain sample, the square of rotation speed increased almost linearly with the decrease of laser wavelength (or increase of photon energy) under the same laser power density. (450, 532 and 650 nm: blue, green and red diamonds; bigger size of the diamond represented higher laser power density.) Error bars in **(d)**, **(e)** and **(f)** are the variance $S^2$ of rotation speed.

**Mechanism of the macroscopic and direct light propulsion**

Before investigation and discussion of the possible mechanism of this surprising bulk scale and direct light manipulation, the composition (Supplementary Table 1 and Supplementary Figs 5–7), morphology and structure (Supplementary Figs 8–10) of the graphene sponge were thoroughly investigated. Based on these results and reports elsewhere[21-23], the graphene sponge should be a 3D cross-linked monolithic graphene material, where the graphene sheets, as the building unit, are covalently cross-linked together through the reactions between the oxygenic functional groups located mostly on the sheets edges during the solvothermal process. The C–O covalent bonds mainly located at the graphene sheet edges not only hold the whole material as a bulk and monolithic object structurally, but also act as an electronic barrier and induce quantum confinement between the graphene sheets electronically through a localized band gap[24-26]. Therefore, each of the graphene sheets or sp$^2$ domains in the bulk material can be thought as an electronically isolated and structurally suspended individual graphene sheet. So overall, the graphene sponge can be seen and treated as a sum of many individual graphene sheets electronically, but without exhibiting the strong-coupling properties of the graphene sheets as in the case of graphite. Thus, the Dirac type band structure should be essentially maintained for the individual graphene sheets in the graphene sponge except with a slightly opened band gap, which should allow the material to behave as a collection of individual suspended graphene sheets with the intrinsic properties of graphene retained[17,27,28]. Note the photon energy of the lasers that we used is at least ~1.91 eV, which should be much higher than the possibly very small band gap[29,30] of the individual graphene sheets in the graphene sponge.

Two working mechanisms have been well documented for beam-powered propulsion[13]: either an external laser beam ablates/burns off propellant to provide propulsion similar to conventional chemical rockets[12,13], or the direct radiation pressure generates the propulsion force governed by the Maxwell electromagnetism theory as has been proposed for the solar sail[14,15]. The light intensities (irradiance) of Watt level laser and simulated sunlight in our tests were at $10^5$ and $10^4$ W m$^{-2}$ level respectively. Based on the radiation pressure theory, the propulsion forces produced by the radiation pressure of such laser and simulated sunlight should be both at ~$10^{-9}$ N and they are orders of magnitude smaller than the force required to move and propel the bulk graphene object (detailed in SI). So the direct radiation pressure induced mechanism can be excluded. Another possibility for explaining our laser-induced propulsion and rotation is the conventional laser beam ablating or burning off of graphene material to generate a plasma plume or carbon particles and molecules for propulsion. But such a mechanism normally needs extremely high laser power supply, so pulsed laser sources (ms/ns level pulse width and gigawatt level peak power) or ultrahigh power continuous wave laser (up to megawatt level) were used[13]. This is contrary to our light-induced motion which can even be observed with sun light which has a much lower power. Note that the continuous wave lasers that we used were only at the Watt level. Furthermore, all the experimental results we have so far as discussed below exclude such a mechanism. Firstly, for all the graphene sponge samples after repeated testing, no noticeable ablation or combustion trace was observed and there was no evidence for weight reduction (detailed in SI). Secondly, no detectable carbon clusters or other small molecular pieces/particles from the graphene sponge under our Watt level continuous wave laser illumination was observed using even high resolution mass spectrometers in the mass/charge range of 12-4000 (Supplementary Figs 11 and 12). Lastly, it should be noted that graphene or graphite can withstand high temperature without decomposing[22] and the fabrication of our material involves annealing at 800 ℃ in the last step. Based on these observations, we believe that the direct laser ablation is unlikely to provide the main driving force in our experiments.

These results prompt us to search for other possible mechanisms for macroscopic direct light manipulation. It is well known that graphene sheet shows unique optoelectronic properties due to its Dirac conical and gapless band structure, which allows graphene to: 1) absorb all wavelength of light efficiently, 2) achieve population inversion state easily as a result of the excitation of hot electrons and the relaxation bottleneck at the Dirac point and then 3) eject the hot electrons following the Auger-like mechanism[16-19,31-33]. Many studies of this effect have been reported not only for individual suspended graphene sheets[17,18] but also for reduced graphene oxide sheets[29]. In the competition of different relaxation pathways of carriers at the reverse saturated state of the optically excited graphene, due to the weak electron-phonon coupling, the Auger-like recombination is proved to be the dominant process and plays an unusually strong role in the relaxation dynamics process[20,33-35] of the hot carriers (electrons). It has also been reported that the fully-suspended graphene shows much enhanced photoresponsivity than graphene on a support[17,36]. As discussed above, our bulk graphene material could be treated as a macroscopic collection of many electronically isolated individual graphene sheets, thus a bulk scale sum of such a photoresponse should be observed due to the macroscopic addition of many individual and suspended graphene sheets in this unique graphene material. So, with continuous laser/light excitation, long-lived photoexcited and energetic hot carriers (electrons) would be excited into the conduction band and a population inversion could be generated and maintained. Note that such a distribution of hot carriers could be obtained by two ways: either by increasing the absorbed photon density (laser power density) or by increasing the photon energy (laser frequency), which are completely interchangeable[37] and is supported by our results. Thus, we argue that the Auger-like recombination is probably also the dominant path for the relaxation of the hot electrons for our photoexcited graphene, and if it involves the inner energy levels as in the classical Auger effect, will result in the hot carriers (electrons) being ejected out as free electrons after they obtain enough energy (Fig. 3a)[33,35,37]. Note the ejected electrons will be emitted randomly in all directions, and some will be absorbed by the surrounding graphene sponge and some will generate a mutually offsetting force, only the net electrons ejected in the direction opposite of the laser beam propagation direction can contribute to the net propulsion and push the sample in the direction of laser beam (Fig. 3b).

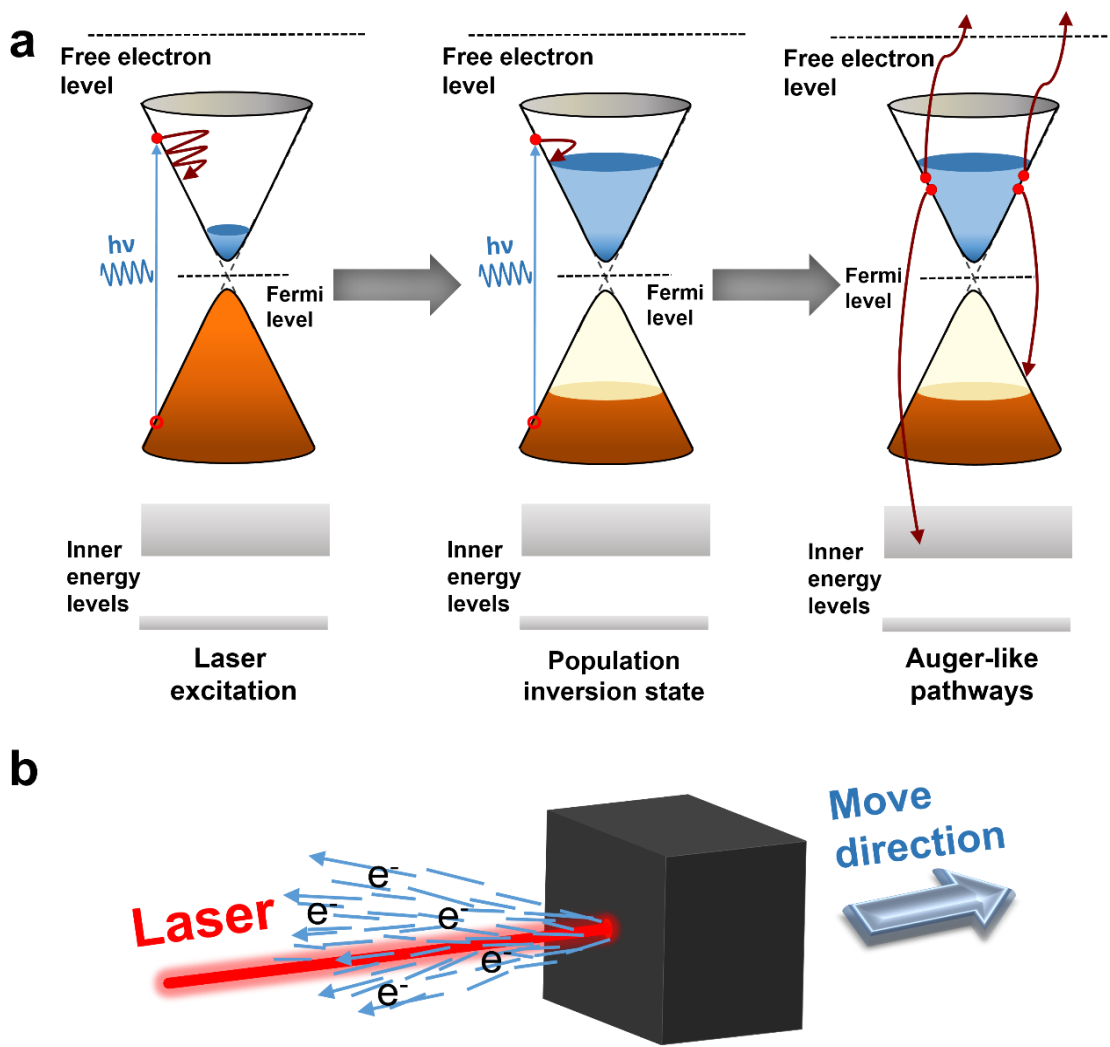

**Figure 3 | Schematic diagrams of the proposed mechanism. a,** Schematic of the proposed mechanism of electron emission: laser excites electrons from the valence

band to conduction band and a population inversion state is achieved and maintained, some hot electrons obtained enough energy to be ejected out and to become free electrons through Auger-like pathways. **b,** Schematic diagram showing the net emitted electrons flying away from the graphene sponge and propelling the graphene object along the laser propagation direction.

**Measurement of ejected electrons**

To verify the mechanism we designed a device to collect the ejected electrons from the graphene sponge under laser illumination (Fig. 4a), where the sample was put inside a metal box (as the electron collecting electrode) which was put into a vacuum chamber with a quartz window for laser illumination. With a 450 nm and 1.5 W chopped continuous wave laser, a strong cycling current signal appeared immediately once the laser was illuminated on the graphene sponge and matched the chopped laser beam cycling (Fig. 4b). Similar results were also observed using lasers with other wavelength (Supplementary Fig. 13). More results for current signal intensity with different wavelengths and power densities are shown in Fig. 4c (also Supplementary Figs 13 and 14), where the average current due to the electron ejection rate has a linear relationship with laser power density for the same laser wavelength and also has a linear relationship with laser wavelength (or photon energy) for the same laser power density. The energy of the emitted electrons was further measured using a home-modified X-ray Photoelectron Spectroscopy (XPS), and the result is shown in Fig. 4d, where a broad kinetic energy distribution was observed for the ejected electrons. We also measured a series of control materials including carbon nanotubes by the same device under the same conditions, but neglectable current signals were obtained (Supplementary Fig. 15).

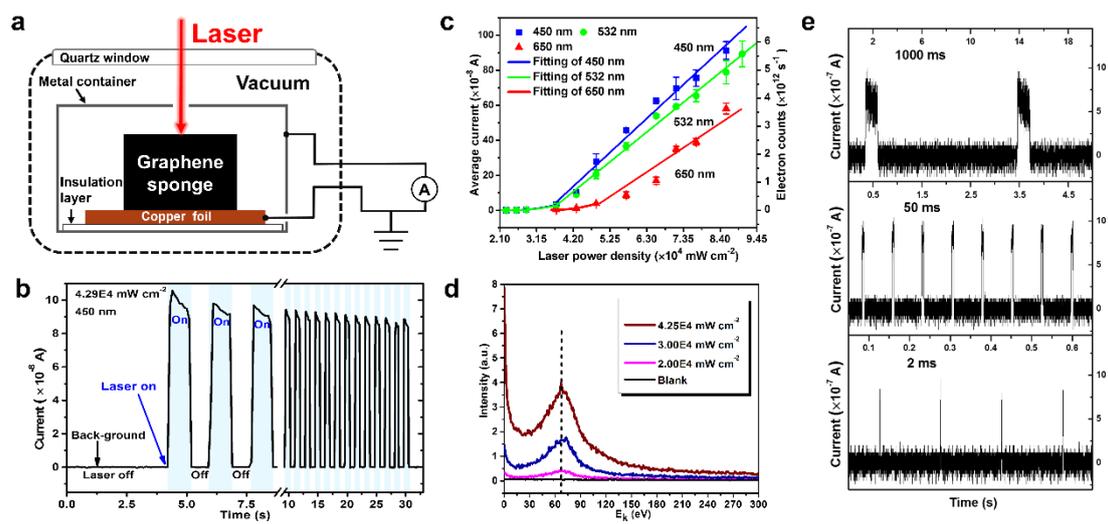

**Figure 4 │ Measurement of the electron emission of graphene sponge under laser illumination. a**, Schematic of the device for measuring electrons emitted from the sample. **b**, A typical curve obtained by measuring the current intensity and the on-off state of the laser was controlled with a chopper. **c**, The average current signal intensity could be obtained (detailed in SI), and for a given laser wavelength, the intensity increased linearly with the laser power density in a wide range. The error bars represented Standard Deviation (SD) for the repeated same measurement. **d**, Kinetic energy distribution spectrum of electrons emitted from graphene sponge under laser (450 nm) illumination showed a broad energy distribution. **e**, Current signals under the illumination of laser pulse with different pulse widths (1000, 50 and 2 ms), and neither time-related delay impact nor meaningful current intensity change was observed. The slight difference between different signals should be caused by measurement error.

It is possible that the electron emission is due to the conventional thermionic mechanism (Edison effect) following the Richardson equation ($J = A_G T^2 \exp(-w/kT)$), which depends on the temperature exponentially[38]. If this is the case, temperature effects should be observed - the heating/cooling process is much slower than the rather fast photonic process. But as shown in Fig. 4e and Supplementary Fig. 16 for the plot of in situ (real time) time versus current, both the current intensity and pattern does not show such time-related effect for all the tests with different laser pulse widths in a wide range from 1000 to 2 ms. Furthermore, the estimation shows that the temperature of the graphene sponge could not be higher than 900 ℃ under our laser pulse illumination even assuming all the laser energy was converted to heat (detailed in SI) without any energy loss. Note generally, efficient thermionic emission temperature is at least 1000 ℃ for most materials[39], and even higher for carbon nanotube and graphene[40,41]. Lastly, as shown in Fig. 4c and Supplementary Fig. 14, under the same laser power density, the current signal intensity had a clear wavelength dependence. With the well-known fact that graphene has efficient absorption over the full spectrum (also see Supplementary Fig. 17), the above results indicate that at least the thermionic mechanism should not be the major path[40]. So above all, the electron emission of the graphene sponge under laser illumination should be a direct photo-induced process essentially.

With all these experimental results, the remaining question is whether the kinetic energy generated by the ejected electrons is large enough to move/propel the sample. The average current was measured at about $3.0 \times 10^{-8}$-$9.0 \times 10^{-7}$ A under the laser power 1.3-3.0 W (450 nm, power density $3.71 \times 10^{4}$-$8.57 \times 10^{4}$ mW cm$^{-2}$ for 3.5 mm$^2$ laser spot, Fig. 4c), which means that the electron ejection rate should be about $2.0 \times 10^{11}$-$5.7 \times 10^{12}$ s$^{-1}$, so a power of $2.2 \times 10^{-6}$-$6.4 \times 10^{-5}$ J s$^{-1}$ (Watt) could be obtained based the average kinetic energy (Fig. 4d) of 70 eV for the ejected electrons. This is larger than the energy necessary ($< 10^{-6}$ Watt) to vertically propel the sample (detailed in SI). Rotation is easier to achieve, compared with the laser-induced vertical propulsion. Note the actual propulsion force/energy should be significantly larger than the values estimated above, since clearly not all the electrons were collected in the measurement. Thus, this propulsion by Light-Induced Ejected Electrons (LIEE) is actually an energy transfer process, where the photon energy is absorbed by graphene bulk materials and converted into the kinetic energy of ejected electrons, rather than a direct momentum transfer process like in the earlier proposed propulsion by light pressure. In the light of the complicated relaxation process for the light-induced hot electrons at the reverse saturated state, much more works, including a comprehensive theoretical modeling, are needed to fully understand this novel phenomenon.

**Discussion**

It is important to emphasize that the remarkable light-induced macroscale propulsion reported herein is a result of the unique electronic band structure at the Dirac point and associated optoelectronic properties of the graphene sheet itself together with the unique macro structural character of this novel bulk graphene material. Obviously, other 2D materials with similar Dirac conical band structure such as graphynes[42], silicene[43], planar Ge[44] and 2D $Bi_{1-x}Sb_x$ thin films[45], if assembled in a similar way, might show the similar LIEE phenomena when illuminated with light. Thus some macroscopic practical utilization of the optical force, only observed for the microscale light actuation to date[2,8], may be achieved based on this work. Furthermore, in this process, the propulsion is generated by the ejected electrons, which is completely different from the conventional laser ablation propulsion. While the propulsion energy/force is still smaller compared with conventional chemical rockets, it is already several orders larger than that from light pressure. Assuming the area of a typical solar-cell panel structure on the satellite is ~50 $m^2$ and because a laser-graphene sponge-based rocket does not need other moving parts, with a payload of 500 kg, the acceleration rate would be 0.09 m $s^{-2}$. Since the density of graphene sponge is very low and no other onboard propellant is needed (the required vacuum and light are naturally available in space), the theoretical specific impulse of our laser propulsion could be much higher. Furthermore, the material could also be used as a novel and convenient electron emission source.

In summary, our results demonstrate that macro graphene based objects could be propelled by a Watt level laser and even sunlight directly up to the sub-meter scale following a novel LIEE mechanism. The propulsion could be further enhanced by increasing the light intensity and/or improving the illumination area. For example, using an adjustable laser array, the force needed for attitude control and orbital adjustment of a spacecraft, and even transporting a payload in outer space could be achieved using light directly. Other 2D materials in addition to graphene with the Dirac conical band structure are also expected to demonstrate this striking property. These results also indicate that exotic and unprecedented properties or phenomena could be obtained when these unique 2D materials are assembled in such a way where their intrinsic 2D properties are retained.

**Method**

The synthesis of graphene sponge and the preparations of the samples for experiments and characterization were shown in SI. The graphene sponge samples in Figs 2a and 2b were placed in the vertical vacuum tube for observing the laser-induced propulsion, and the vacuum was $6.8 \times 10^{-4}$ Torr. The laser spot areas for Figs 2a and 2b were all ~4 mm$^2$. The diameter and height of the cylinder shape sample were 10 and 11 mm respectively, and the mass of the sample was ~0.86 mg. Two different graphene sponge samples were used in Fig. 2e and recoded as sample A and B. The samples in Fig. 2c, 2d and 2f were all sample A in Fig. 2e, and sample A had a size of $12 \times 7 \times 5$ mm$^3$ and a weight of 0.44 mg. Sample B in Fig. 2e had a size of $12.5 \times 8 \times 3.5$ mm$^3$ and a weight of 0.36 mg. The laser spot areas in Figs 2c-2f were all about 4.5 mm$^2$ and all the experiments were performed in vacuum environment of $6.8 \times 10^{-4}$ Torr. The laser spot areas in Fig. 4 were all about 3.5 mm$^2$. In Fig. 4a, the distance between the sample and the metal container as the current collection electrode is ~2 mm, and the vacuum was better than $5 \times 10^{-6}$ Torr. In Fig. 4d, electron kinetic energy was measured by the Concentric Hemispherical Electron Energy Analyzer (CHA) on a XPS instrument with slight modification, and graphene sponge was illuminated with laser but without X-ray radiation. The vacuum of the XPS instrument was better than $6.7 \times 10^{-9}$ Torr. Measurement detail was supplied in SI. In Fig. 4e, the laser wavelength was 450 nm and the power density was ~$8.57 \times 10^4$ mW cm$^{-2}$. A digital oscilloscope with high enough sampling frequency was used to record the current signals in real time.

# References

1. Ashkin, A. Acceleration and trapping of particles by radiation pressure. *Phys. Rev. Lett.* **24**, 156-159 (1970).

2. Grier, D. G. A revolution in optical manipulation. *Nature* **424**, 810-816 (2003).

3. Swartzlander, G. A., Peterson, T. J., Artusio-Glimpse, A. B. & Raisanen, A. D. Stable optical lift. *Nature Photonics* **5**, 48-51 (2011).

4. Dogariu, A., Sukhov, S. & Saenz, J. J. Optically induced 'negative forces'. *Nature Photonics* **7**, 24-27 (2013).

5. Kane, B. Levitated spinning graphene flakes in an electric quadrupole ion trap. *Phys. Rev. B* **82**, 115441 (2010).

6. Marago, O. M. *et al.* Brownian motion of graphene. *ACS Nano* **4**, 7515-7523 (2010).

7. Twombly, C. W., Evans, J. S. & Smalyukh, I. I. Optical manipulation of self-aligned graphene flakes in liquid crystals. *Opt. Express* **21**, 1324-1334 (2013).

8. Ashkin, A. History of optical trapping and manipulation of small-neutral particle, atoms, and molecules. *IEEE J. Sel. Top. Quantum Electron.* **6**, 841-856 (2000).

9. Shvedov, V. G. *et al.* Giant optical manipulation. *Phys. Rev. Lett.* **105**, 118103 (2010).

10. Shvedov, V. G., Hnatovsky, C., Rode, A. V. & Krolikowski, W. Robust trapping and manipulation of airborne particles with a bottle beam. *Opt. Express* **19**, 17350-17356 (2011).

11. Kobayashi, M. & Abe, J. Optical motion control of maglev graphite. *J. Am. Chem. Soc.* **134**, 20593-20596 (2012).

12. Ageev, V. P. *et al.* Experimental and theoretical modeling of laser propulsion. *Acta Astronaut.* **7**, 79-90 (1980).

13. Phipps, C. *et al.* Review: laser-ablation propulsion. *J. Propul. Power* **26**, 609-637 (2010).

14. Tsu, T. C. Interplanetary travel by solar sail. *ARS J.* **29**, 422-427 (1959).

15. Tsuda, Y. *et al.* Flight status of IKAROS deep space solar sail demonstrator. *Acta Astronaut.* **69**, 833-840 (2011).

16. Nair, R. R. *et al.* Fine structure constant defines visual transparency of graphene. *Science* **320**, 1308 (2008).

17. Patil, V., Capone, A., Strauf, S. & Yang, E. H. Improved photoresponse with enhanced photoelectric contribution in fully suspended graphene photodetectors. *Sci Rep* **3**, 2791 (2013).

18. Li, T. *et al.* Femtosecond population inversion and stimulated emission of dense Dirac fermions in graphene. *Phys. Rev. Lett.* **108**, 167401 (2012).

19. Perakis, I. E. Stimulated near-infrared light emission in graphene. *Physics* **5**, 43 (2012).

20. Strait, J. H. *et al.* Very slow cooling dynamics of photoexcited carriers in graphene observed by optical-pump terahertz-probe spectroscopy. *Nano Lett.* **11**, 4902-4906 (2011).

21. Wu, Y. *et al.* Three-dimensionally bonded spongy graphene material with super compressive elasticity and near-zero Poisson's ratio. *Nature Commun.* **6**, 6141 (2015).

22. Xu, Y., Sheng, K., Li, C. & Shi, G. Self-assembled graphene hydrogel via a one-step hydrothermal process. *ACS Nano* **4**, 4324-4330 (2010).

23. Zhou, Y., Bao, Q., Tang, L. A. L., Zhong, Y. & Loh, K. P. Hydrothermal dehydration for the "green" reduction of exfoliated graphene oxide to graphene and demonstration of tunable optical limiting properties. *Chem. Mater.* **21**, 2950-2956 (2009).

24. Wehling, T. O., Katsnelson, M. I. & Lichtenstein, A. I. Impurities on graphene: Midgap states and migration barriers. *Phys. Rev. B* **80**, 085428 (2009).

25. Wu, X. *et al.* Epitaxial-graphene/graphene-oxide junction: an essential step towards epitaxial graphene electronics. *Phys. Rev. Lett.* **101**, 026801 (2008).

26. Loh, K. P., Bao, Q. L., Eda, G. & Chhowalla, M. Graphene oxide as a chemically tunable platform for optical applications. *Nature chem.* **2**, 1015-1024 (2010).

27. Meyer, J. C. *et al.* The structure of suspended graphene sheets. *Nature* **446**, 60-63 (2007).

28. Prechtel, L. *et al.* Time-resolved ultrafast photocurrents and terahertz generation in freely suspended graphene. *Nature Commun.* **3**, 646 (2012).

29. Kim, J. *et al.* Unconventional terahertz carrier relaxation in graphene oxide: observation of enhanced auger recombination due to defect saturation. *ACS Nano* **8**, 2486-2494 (2014).

30. Eda, G., Mattevi, C., Yamaguchi, H., Kim, H. & Chhowalla, M. Insulator to semimetal transition in graphene oxide. *J. Phy. Chem. C.* **113**, 15768-15771 (2009).

31. Winzer, T., Knorr, A. & Malic, E. Carrier multiplication in graphene. *Nano Lett.* **10**, 4839-4843 (2010).

32. Winzer, T. & Malic, E. Impact of Auger processes on carrier dynamics in graphene. *Phys. Rev. B* **85**, 241404 (2012).

33. Winzer, T., Malic, E. & Knorr, A. Microscopic mechanism for transient population inversion and optical gain in graphene. *Phys. Rev. B* **87**, 165413 (2013).

34. Brida, D. *et al.* Ultrafast collinear scattering and carrier multiplication in graphene. *Nature Commun.* **4**, 1987 (2013).

35. Gabor, N. M. Impact excitation and electron-hole multiplication in graphene and carbon nanotubes. *Acc. Chem. Res.* **46**, 1348-1357 (2013).

36. Gabor, N. M. *et al.* Hot carrier-assisted intrinsic photoresponse in graphene. *Science* **334**, 648-652 (2011).

37. Tielrooij, K. J. *et al.* Photoexcitation cascade and multiple hot-carrier generation in graphene. *Nature Phys.* **9**, 248-252 (2013).

38. Dushman, S. Electron emission from metals as a function of temperature. *Phys. Rev.* **21**, 623-636 (1923).

39. Turner, L. W. (ed) *Electronics engineer's reference book (Fourth Edition).* (Newnes-Butterworth, London, 1976).

40. Yaghoobi, P., Moghaddam, M. V. & Nojeh, A. "Heat trap": Light-induced localized heating and thermionic electron emission from carbon nanotube arrays. *Solid State Commun.* **151**, 1105-1108 (2011).

41. Oida, S., Hannon, J. B., Tromp, R. M., McFeely, F. R. & Yurkas, J. A simple in situ method to detect graphene formation at SiC surfaces. *Appl. Phys. Lett.* **98**, 213106 (2011).

42. Malko, D., Neiss, C., Vines, F. & Gorling, A. Competition for graphene: graphynes with direction-dependent Dirac cones. *Phys. Rev. Lett.* **108**, 086804 (2012).

43. Vogt, P. *et al.* Silicene: compelling experimental evidence for graphenelike two-dimensional silicon. *Phys. Rev. Lett.* **108**, 155501 (2012).

44. Cahangirov, S., Topsakal, M., Akturk, E., Sahin, H. & Ciraci, S. Two- and one-dimensional honeycomb structures of silicon and germanium. *Phys. Rev. Lett.* **102**, 236804 (2009).

45. Tang, S. & Dresselhaus, M. S. Constructing anisotropic single-Dirac-cones in $Bi_{1-x}Sb_x$ thin films. *Nano Lett.* **12**, 2021-2026 (2012).

## Acknowledgements

The authors gratefully acknowledge financial support from the MoST (Grants 2012CB933401 and 2014CB643502), NSFC (Grants 91433101, 21374050 and 51273093) and PCSIRT (IRT1257). The authors thank for the help in XPS from Dr. Zhanping Li (Tsinghua University) and mass spectrum measurements from Prof. Xianglei Kong (Nankai University) and Prof. Haiyang Li (Dalian Institute of Chemical Physics, Chinese Academy of Sciences).

**Author Contributions**

Y.C. conceived and directed the study. T.Z. and H.C. carried out most of the experiments and data analysis, with some initial experiments was done by Y.W.. T.Z., and Y.C. together with H.C. prepared most of the manuscipt. H.C. synthesised most of the samples and prepared the Videos. Y.W., P.X., N.Y. and Y.L. participated in some experiments, data analysis and discussions. K.Z., X.Y. and Z.L participated the current measurermnt. All authors participated in the project discussions and production of the final manuscript.

**Additional information**

Supplementary information is available in the online version of the paper. Reprints and permissions information is available online at www.nature.com/reprints. Correspondence and requests for materials should be addressed to Y. C.

**Competing financial interests**

The authors declare no competing financial interests.